\documentclass{article}
\usepackage{spconf,amsmath,graphicx,hyperref, bm, xcolor}
\usepackage{caption}
\usepackage{subcaption}
\usepackage{multirow}

\newif\ifmk
\mktrue

\title{Rethinking Magnitude Compression in Spectral Losses\\for DDSP Parameter Estimation}
\name{Byunghoo Park \qquad Jayeon Yi \qquad Minje Kim\thanks{The authors appreciate Amazon for providing AWS credits through the Amazon AI PhD Fellowship program.}}
\address{University of Illinois Urbana-Champaign, Siebel School of Computing and Data Science, USA 61801}
\begin{document}
\ninept
\maketitle
\begin{abstract}
Spectral losses for differentiable digital signal processing (DDSP) commonly compress magnitudes, for example by using log-magnitude terms. Such compression mimics human perception, but whether it helps the optimization is less clear, and is usually found only by comparing trained models. In this work, we show that whether compression helps depends on the system's informative region: the region of its dynamic range that carries the information pointing toward the target. To indicate where this region lies without having to train, we introduce \emph{concordance}, which measures how well a loss orders candidate parameter settings by their distance from the target. We then compare uncompressed, compressed, and combined losses by fitting a modal plate reverb and a subtractive synthesizer model to target sounds. A third system, selected for its low-energy informative region, tests the relationship in the opposite direction. In every system, the loss that matches the informative region performs best: the uncompressed one where the region lies at high energies, and the compressed one where it lies low. Surprisingly, the common combination of both log-magnitude and linear-magnitude terms performs best in none of the systems. Code and audio examples are available online\footnote{\url{https://github.com/superstem000/compression-analysis}}.
\end{abstract}

\section{Introduction}
\label{sec:intro}

Reproducing a target sound with a known signal model is a long-standing problem in audio, from synthesizer sound matching to physical-model fitting. Differentiable digital signal processing (DDSP) has made fitting such a model practical by backpropagating loss gradients through its operations, yielding a compact and interpretable reproduction and a domain-appropriate inductive bias \cite{engel2020ddsp}.

Unlike unconstrained waveform generation, optimization in such systems acts through a relatively small set of parameters. Changing these parameters can produce structured and potentially large changes in the spectrum, such as moving spectral peaks or changing resonances and filter responses. Consequently, selecting a loss that provides informative gradients with respect to these parameters, while remaining aligned with the goal of the task, is particularly important \cite{caspe2022}. However, no consensus has been reached on the best spectral loss configuration across tasks \cite{hayes2024}. The optimal loss also varies across signal models, and loss selection remains under-explored \cite{salimi2025}.

One particularly consequential component of loss selection is magnitude compression, which increases the relative contribution of low-energy components. Audio magnitudes span a large dynamic range, whereas perceived loudness grows sublinearly with physical sound intensity \cite{StevensS1955loudness, StevensS1957psychophysical}. Hence, logarithmic and power-law representations have long been used to compress this dynamic range when representing or comparing audio \cite{SmithB1957mu-law, OordA2016wavenet, hayes2024}. A primary example is the multi-scale spectral (MSS) loss, which combines linear-magnitude and log-magnitude terms across multiple time-frequency resolutions \cite{engel2020ddsp, yamamoto2020} and has been widely adopted in DDSP applications.

Although compression is well motivated for perceptual representations, its effect in DDSP parameter estimation is less obvious. Compression can be advantageous when useful information is carried by weak spectral components. In a parametric DDSP system, however, parameter changes may have their most pronounced effects on a small number of dominant spectral components, in which case compression can suppress the gradients most useful for recovering the target parameters. Thus, the perceptually motivated preference for compressed magnitudes need not coincide with the representation that is most informative for fitting a signal model.

On the other hand, an uncompressed linear-magnitude loss is rarely used as a primary objective and instead appears more often as a point of comparison \cite{salimi2025, hayes2023}. Prior works have examined the effect of compression in specific settings. For the MSS loss, compression has been found to exacerbate spectral leakage, with a lesser degree of compression proposed as a remedy \cite{schwar2023}. In speech enhancement, among magnitude-domain losses, a loss with power-law compression, which is milder than a logarithm, was found to perform on par with an uncompressed $L_1$ distance, and both outperformed a log-compressed loss \cite{braun2020}. In speech dereverberation, an intermediate power exponent performed best, which the authors attributed to over-compression of the high-energy regions that carry formant information \cite{li2021}. Each of these explanations, however, is specific to its system. To our knowledge, no general criterion has been established that identifies, for a given system and prior to empirical loss comparison, which portions of its dynamic range carry the information needed to optimize its parameters.

In this paper, we examine magnitude compression as a component of loss selection where the choice can be informed rather than found by comparison. We show that whether compression benefits a system depends on which region of its dynamic range carries information that indicates the direction toward the target, a region we call the \emph{informative region}. We also introduce a measure, \emph{concordance}, that indicates where the informative region lies and aligns with the observed outcomes. Specifically, a system whose informative region lies at high energies is better optimized by an uncompressed loss, and one whose informative region lies at low energies by a compressed one. To illustrate this relationship, we compare linear, log, and hybrid variants, obtained by removing either term from each system's baseline loss. We evaluate them on a modal plate physical model \cite{gabrielli2026} and a subtractive synthesizer \cite{masuda2021}, against ground-truth parameters on rendered targets and perceptually on real recordings. Concordance indicates a high-energy informative region in both, so we also examine a third system whose informative region lies at low energies, to test the relationship in the opposite direction.

\section{Problem Definition}
\label{sec:definitions}

We consider estimating parameters for a differentiable signal model. Such a model $\mathcal{F}$ renders audio $\hat{y} = \mathcal{F}(x; \bm\theta)$ from a parameter setting $\bm\theta$ and any conditioning input $x$. Given a target audio signal $y$, the task is to recover a parameter setting $\bm\theta$ whose rendering $\hat{y}$ resembles it. This can be done by training a network $g_\phi$ that maps a target signal $y$ to a parameter setting. The network is trained on a dataset of rendered audio and their parameters. One straightforward way to define the optimization task is to compare the ground-truth parameters $\bm\theta$ with the network's estimate $\hat{\bm\theta} = g_\phi(y)$. The other is to compare $y$ with its re-rendering $\hat{y} = \mathcal{F}(x; g_\phi(y))$ in the signal space, which is possible because $\mathcal{F}$ is differentiable. We call these two cases the \emph{parameter} and \emph{spectral} losses, respectively. For the latter, we use a time-frequency representation as in Eq.~\ref{eq:loss} instead of the time-domain audio, hence the name \textit{spectral}. This work examines magnitude compression in the spectral loss.

\subsection{Systems}
\label{ssec:systems}

\vspace{0.4em}\noindent\textbf{Modal plate.} The first system is the plate reverb simulator of the 1st DAFx parameter estimation challenge \cite{gabrielli2026}, which implements the general solution of the damped Kirchhoff--Love equation and is described in full there. The challenge varies a fixed subset of the simulator's physical parameters and holds the rest constant, including damping.

Since we evaluate against measured impulse responses of an EMT 140 plate reverb, we use a parameter subset and ranges that differ from the challenge's, chosen to better represent that plate. The estimated parameters are $\{h, \rho, L_y, T_0, \tau_0, r\}$: the thickness, volumetric density, height, tension, decay time at DC, and the decay ratio $r := \tau_1/\tau_0$, where $\tau_1$ is the decay time at a second reference frequency. We set this second reference frequency to 10\,kHz, rather than the challenge's 500\,Hz, to avoid extrapolating the decay of all frequencies above 500\,Hz. Estimating the ratio $r$ instead of $\tau_1$ enforces the passivity condition $\tau_0 > \tau_1$ that the model requires. Impulse responses are rendered at 44.1\,kHz for 1.0\,s. Parameter bounds and fixed values are given in the accompanying repository.

\vspace{0.4em}\noindent\textbf{Subtractive synthesizer.} The second system is the differentiable synthesizer of Masuda and Saito \cite{masuda2021}, which models a two-oscillator subtractive synthesizer with a resonant filter. Table~\ref{tab:synthparams} lists the estimated parameters in the first column. Unlike the original system, which estimates the fundamental, we supply it. This reduces frequency estimation to a single free parameter, the ratio $\kappa$ of the second oscillator's frequency to the first, making frequency comparisons more tractable and perceptual comparisons easier. The real recordings we use are labeled with their fundamental, so supplying it reflects information available in practice.

\begin{table}[t]
\centering
\caption{Estimated parameters of the synthesizer and reverb systems. Numbers give how many of each parameter are estimated; amplitude and oscillator mix are estimated per oscillator.}
\label{tab:synthparams}
\begin{tabular}{llcc}
\hline
 & & Synth. & Reverb \\
\hline
$a$       & amplitude          & 2  & -- \\
$f_c$     & filter cutoff      & 1  & 1 \\
$f_0$     & fundamental        & -- & 1 \\
$\kappa$  & frequency ratio    & 1  & -- \\
$m$       & oscillator mix     & 2  & 1 \\
$Q$       & filter resonance   & 1  & 1 \\
$g$       & reverb gain        & -- & 1 \\
$\lambda$ & reverb decay rate  & -- & 1 \\
\hline
          & total              & 7  & 6 \\
\hline
\end{tabular}
\end{table}

\vspace{0.4em}\noindent\textbf{Reverb system.} The third system uses the same synthesizer as the subtractive synthesizer with a single oscillator, followed by a reverb module. The module convolves the signal with an exponentially decaying impulse response and adds the result to the dry signal. The gain $g$ scales the impulse response and $\lambda$ sets its decay rate. The estimated parameters are listed in the second column of Table~\ref{tab:synthparams}. The system was selected because the concordance analysis of Section~\ref{sec:results_conc} indicated a low-energy informative region, so it tests the relationship in the opposite direction.

\subsection{Loss Variants}
\label{ssec:losses}

We take each system's established loss as the baseline and vary magnitude compression. For the plate, the baseline is the $L_1$ distance on a single magnitude spectrogram, as used in our previous work \cite{park2026}. For the synthesizer and reverb systems, it is the MSS loss of Masuda and Saito \cite{masuda2021} over six resolutions. Both losses take the form
\begin{equation}
\label{eq:loss}
\begin{split}
\mathcal{L} = \frac{1}{N}\sum_{i=1}^{N} \Big[
  &\alpha_{\mathrm{lin}}\, \overline{\big|S_i(y) - S_i(\hat{y})\big|} \\
  + &\alpha_{\mathrm{log}}\, \overline{\big|\log(S_i(y){+}\epsilon)
      - \log(S_i(\hat{y}){+}\epsilon)\big|} \Big],
\end{split}
\end{equation}
where $N$ is the number of resolutions, $y$ the target and $\hat{y}$ the rendering, $S_i(\cdot) = |\mathrm{STFT}_i(\cdot)|$ the magnitude spectrogram at the $i$-th resolution, computed with a Hann window at 75\% overlap, and $\overline{(\cdot)}$ the mean over bins and frames. The constant $\epsilon$ is taken from each reference implementation: $10^{-7}$ for the plate \cite{gabrielli2026} and $10^{-2}$ for the synthesizer \cite{masuda2021}. Each $\alpha$ is either set to zero or kept at its baseline value, giving three variants per system that differ only in compression: \emph{linear} ($\alpha_{\mathrm{log}} = 0$), \emph{log} ($\alpha_{\mathrm{lin}} = 0$), and \emph{hybrid} (both nonzero).

\section{Loss Alignment Analysis}
\label{sec:measure}

We analyze the loss terms directly, examining how each ranks rendered candidates as a system's parameters are varied around a target. The analysis requires no trained model and indicates which magnitude weighting aligns with the system's informative region. Its results are given in Section~\ref{sec:results_conc}. An uncompressed and a log-compressed loss differ only in how they weight the dynamic range. The linear term weights by absolute magnitude, so its contribution is concentrated at high energies. The log term weights relative differences equally across the range, which shifts weight from high to low energies compared with the linear term.

Our hypothesis is that a loss is a more reliable optimization signal for a system when its weighting coincides with that system's informative region. A change in a parameter alters the spectrum unevenly across the dynamic range. Since every candidate the model can produce is determined by its parameters, what distinguishes one candidate from another is exactly the sum of each parameter's alteration of the spectrum. Therefore, a loss that places more weight on the regions where these alterations occur may be a more reliable optimization signal. We measure this alignment directly by asking how often each loss orders candidates by their true distance from the target.

For a given target, we draw $K$ candidate parameter settings and render each. Let $d_i$ be the distance between candidate $i$'s parameters and the target's, and $L_i$ the value of a loss $L$ between the rendered candidate and the target. We define the \emph{concordance} as
\begin{equation}
\label{eq:conc}
\mathcal{C}(L) = \Pr\big[\, L_i < L_j \;\big|\; d_i < d_j \,\big],
\end{equation}
the empirical probability, over all ordered pairs, that the loss ranks two candidates in the same order as their true parameter distance. Targets are drawn at random across the entire parameter space, and candidates are drawn by displacing the target in a random direction, by up to $0.15$ in normalized parameter coordinates. We use 100 targets with 32 candidates each. Computing this for the linear and log terms separately gives $\mathcal{C}_{\mathrm{lin}}$ and $\mathcal{C}_{\mathrm{log}}$. Since the two differ only in magnitude weighting, their difference isolates the effect of compression.

Concordance is applied in two ways. Comparing $\mathcal{C}_{\mathrm{lin}}$ with $\mathcal{C}_{\mathrm{log}}$ indicates whether the informative region lies toward high or low energies. We also trace the log concordance as the loss is restricted to bins within $F$\,dB of the target's peak. Increasing $F$ includes progressively lower-energy bins, so the change in concordance shows how including each region affects the ordering of candidates.

\section{Experimental Setup}
\label{sec:setup}

\vspace{0.4em}\noindent\textbf{Training.} All three systems were trained with Adam in three phases: on the parameter loss alone, then a linear crossfade to the spectral loss, then on the spectral loss alone. The variants of each system started from a single shared checkpoint at the end of the parameter-loss phase, and diverged only once the spectral loss was introduced. For the plate, clips of 1.0\,s were rendered at 44.1\,kHz, and the phases ended at steps 2{,}500, 5{,}000 and 10{,}000. The synthesizer and reverb systems followed the procedure of Masuda and Saito \cite{masuda2021}, with clips of 4.0\,s at 16\,kHz and the phases ending at steps 12{,}500, 50{,}000 and 100{,}000. Within each system, the variants shared all training settings and differed only in the loss weights of Eq.~\ref{eq:loss}.

\vspace{0.4em}\noindent\textbf{Estimator.} Both estimators were ordinary architectures, chosen so that the comparison would rest on the loss rather than on the network. The synthesizer and reverb systems used the estimator of Masuda and Saito \cite{masuda2021}. It applies three strided convolutional blocks of width 64 along the mel axis of a log-mel spectrogram within each frame, followed by a single-layer GRU of 512 units over time and a sigmoid-bounded linear head that outputs the normalized parameters. For the plate, we used a convolutional network of five strided blocks of width 32 over the magnitude spectrogram, with a $\tanh$-bounded head that outputs the normalized parameters directly.

\vspace{0.4em}\noindent\textbf{Evaluation.} Since the variants differ only in their training loss, scoring them with any one of those losses would favor the variant trained on it. The evaluation metric was therefore held fixed across variants and kept distinct from every loss being compared. This also excluded close relatives such as log-spectral distance~\cite{gray1976}, which differs from the log term only in its norm. We report the mel-frequency cepstral coefficient (MFCC) distance, which compares mel-scaled cepstral coefficients rather than STFT magnitudes, while retaining the compression appropriate for perceptual comparison. Where ground truth is available, we also report parameter error as the median normalized mean squared error (NMSE) between estimated and true parameters over the test set.

\vspace{0.4em}\noindent\textbf{Real recordings and listening test.} To test whether the ranking of the variants on rendered targets holds when no parameter setting reproduces the target exactly, we evaluated on real recordings. A target the model cannot approximate carries information that the concordance analysis does not characterize, and on such a target the variants would differ in how they fail rather than in how well they fit. We therefore selected recordings that the model is able to represent.

For the plate, we used 12 \textit{measured} EMT 140 impulse responses spanning three tone settings and four damper settings \cite{emt140}. For perceptual evaluation, each was convolved with a dry source: six snare hits \cite{idmt} and six sung excerpts \cite{vocalset}, sources commonly treated with plate reverb \cite{izhaki2008, izotope_plate}. For the synthesizer, we used single notes of a Moog Minitaur from the Modular Samples library. The Minitaur is a two-oscillator analog synthesizer, similar to the model. We used two non-trivial patches that the model can plausibly represent, \texttt{Doubles} \cite{minitaur_doubles} and \texttt{MT Pad} \cite{minitaur_mtpad}, over the pitch range seen during training.

A MUSHRA-like \cite{bs1534} study was conducted to assess the perceptual similarity of each variant's output to the target. Seven audio experts rated, in 12 trials per system, how similar each stimulus was to the reference. The test used webMUSHRA \cite{webmushra}, with all stimuli loudness-matched to $-23$\,LUFS. The low anchor was a lowpass at 1300\,Hz for the plate and 800\,Hz for the synthesizer, rather than the 3.5\,kHz of BS.1534-3, since the standard cutoff was nearly indistinguishable from the reference on this material.

\begin{figure}[t]
\centering
\includegraphics[width=0.6\columnwidth]{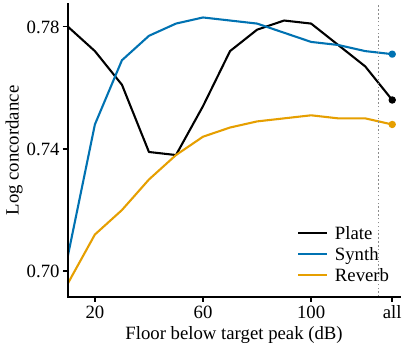}
\caption{Log concordance as the loss is restricted to bins within $F$\,dB of the target's peak. The rightmost point, past the dotted line, is the unrestricted loss.}
\label{fig:floor_sweep}
\end{figure}

\begin{table}[t]
\centering
\caption{Concordance of each loss term over the full spectrum, across 100 targets. The difference $\mathcal{C}_{\mathrm{lin}} - \mathcal{C}_{\mathrm{log}}$ is paired per target and given with its standard error. $t$ is the difference divided by its standard error.}
\label{tab:conc_all}
\begin{tabular}{lcccc}
\hline
System & $\mathcal{C}_{\mathrm{lin}}$ & $\mathcal{C}_{\mathrm{log}}$ & difference & $t$ \\
\hline
Plate    & 0.773 & 0.756 & $+0.017 \pm 0.007$ & $+2.5$ \\
Synth.   & 0.781 & 0.771 & $+0.010 \pm 0.003$ & $+3.5$ \\
Reverb   & 0.711 & 0.748 & $-0.037 \pm 0.005$ & $-7.8$ \\
\hline
\end{tabular}
\end{table}

\begin{table}[t]
\centering
\caption{MFCC distance and parameter NMSE for each loss variant. Lower is better. NMSE is available for rendered targets only.}
\label{tab:results}
\begin{tabular}{llccc}
\hline
 & & \textsc{linear} & \textsc{log} & \textsc{hybrid} \\
\hline
\multirow{3}{*}{Plate}
 & MFCC (rendered) & \textbf{7.88}   & 8.76   & 8.35 \\
 & NMSE            & \textbf{0.036}  & 0.218  & 0.221 \\
 & MFCC (real)     & \textbf{13.92}  & 19.01  & 15.98 \\
\hline
\multirow{3}{*}{Synth.}
 & MFCC (rendered) & \textbf{6.91}   & 11.80  & 12.02 \\
 & NMSE            & \textbf{0.0072} & 0.0157 & 0.0161 \\
 & MFCC (real)     & \textbf{11.45}  & 14.01  & 12.76 \\
\hline
\multirow{2}{*}{Reverb}
 & MFCC (rendered) & 8.63            & \textbf{7.93}   & 9.12 \\
 & NMSE            & 0.0108          & \textbf{0.0049} & 0.0238 \\
\hline
\end{tabular}
\end{table}

\section{Results}
\subsection{Loss Alignment Analysis}
\label{sec:results_conc}

Table~\ref{tab:conc_all} gives the concordance of each term over the full spectrum. For the plate and the synthesizer, the linear term is more concordant, by $+0.017$ and $+0.010$ respectively, indicating a high-energy informative region. For the reverb system, the log term is more concordant, by $0.037$, indicating a low-energy one. All three differences exceed twice their standard error.

Fig.~\ref{fig:floor_sweep} shows the log concordance as the loss is restricted to bins within $F$\,dB of the target's peak. For the two systems where the linear term is more concordant, including the lowest-energy bins lowers the log concordance: the plate falls by $0.026$ from its peak at $90$\,dB to the unrestricted loss, and the synthesizer by $0.012$ from its peak at $60$\,dB. The reverb system falls by only $0.003$ and stays nearly flat from $90$\,dB onward. The bins included last are those the log term up-weights most relative to the linear term. The fall in log concordance as they are included means that these quiet regions are less informative than the louder ones, to the point that including them substantially degrades the ordering of candidates.

\begin{figure}[t]
     \centering
     \begin{subfigure}[b]{0.49\columnwidth}
         \centering
         \includegraphics[width=\textwidth]{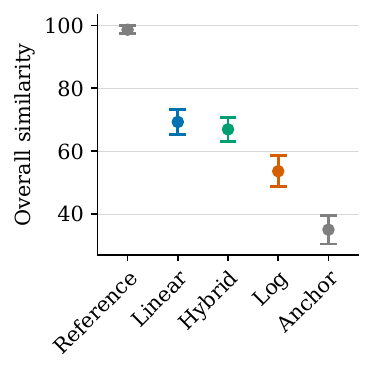}
         \caption{Modal plate}
         \label{fig:mushra_plate}
     \end{subfigure}
     \hfill
     \begin{subfigure}[b]{0.49\columnwidth}
         \centering
         \includegraphics[width=\textwidth]{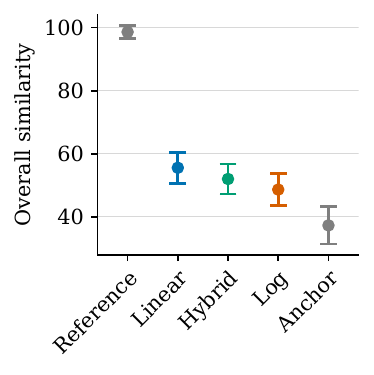}
         \caption{Subtractive synthesizer}
         \label{fig:mushra_synth}
     \end{subfigure}
     \caption{Listening test ratings of overall similarity to the reference, over 12 items and seven listeners. Error bars are 95\% confidence intervals pooled over all ratings.}
     \label{fig:mushra}
\end{figure}

\begin{table}[t]
\centering
\caption{Listening-test contrast in each experiment. Differences are tested with a paired $t$-test on per-listener means over 12 items and seven listeners, two-sided at $\alpha = 0.05$. The last column gives the number of listeners preferring the linear variant.}
\label{tab:mushra}
\begin{tabular}{lcccc}
\hline
System & linear $-$ log & 95\% CI & $p$ & listeners \\
\hline
Plate    & $+15.7$ & $[+4.1,\, +27.3]$ & 0.016 & 6/7 \\
Synth.   & $+6.9$  & $[+0.8,\, +13.1]$ & 0.033 & 6/7 \\
\hline
\end{tabular}
\end{table}

\subsection{Training Results}

\vspace{0.4em}\noindent\textbf{Rendered targets.} Table~\ref{tab:results} gives the objective metrics of the models trained with each loss variant. On targets rendered by the model itself, the results follow the concordance analysis of Section~\ref{sec:results_conc} in every system. For the plate and the synthesizer, the linear variant gives both the lowest parameter error and the lowest MFCC distance. For the reverb system, the log variant does. In all three systems, the two metrics agree on the best variant. The hybrid variant is best in no system. On parameter error it is clearly the worst in the reverb system, and level with the log variant in the other two.

\vspace{0.4em}\noindent\textbf{Real recordings.} On real recordings, MFCC distance ranks the linear variant best, as on rendered targets, in both systems for which recordings were used (Table~\ref{tab:results}). The mean ratings of the listening test give the same ranking (Fig.~\ref{fig:mushra}), but the confidence intervals overlap. Linear and hybrid are not separated in either experiment, and linear and log are separated only on the plate. Compared directly, the linear--log contrast reaches significance in both experiments, comfortably on the plate and by a narrow margin on the synthesizer (Table~\ref{tab:mushra}). Listeners agree on the direction of the difference far more than on its size: six of seven prefer the linear variant in each experiment, but how strongly they prefer it varies widely. Unlike the intervals, the paired contrast bounds the difference directly. The linear variant therefore holds a perceptual advantage over the log variant in both systems, while its comparison with the hybrid is unresolved.

\subsection{Discussion}

Compression moves the loss's weight toward low energies, and whether that helps depends on whether the system's informative region lies there. Fig.~\ref{fig:floor_sweep} shows this directly: the bins the log term up-weights are the last to be included, and including them lowers the log concordance in the plate and the synthesizer but not in the reverb system. This degradation also matches how the compressed variants perform relative to the uncompressed one.

Compression is adopted for its perceptual grounding, and perception remains the criterion here: both MFCC distance and the listening test judge the fit perceptually. However, the plate and synthesizer systems show that a weighting that reflects how loudness is heard is not the weighting that best fits a signal model.

The hybrid variant keeps both terms and is the default across much of the DDSP literature, so it may seem a safe compromise. It is best in none of the three systems, and the linear variant outperforms it on every objective measure in all three. In the reverb system it is worse than either of its components. Combining the two terms is therefore not always constructive: the result need not lie between them, and can be worse than both.

The two applications of concordance agree with each other and with the training results in every system, indicating that they reflect the system rather than the analysis. The reverb system was selected because its concordance indicated a low-energy informative region before any variant was trained, and its trained variants showed the predicted reversal. This suggests concordance may be useful for making the choice ahead of training. Compression is thus a component of loss selection that can be informed by the system, rather than found only by comparison.

\section{Conclusion}
\label{sec:conclusion}

We isolated magnitude compression in three differentiable signal models and introduced concordance, a training-free measure of how well each loss term orders candidates by their distance from the target. Concordance agreed with the training results in every case, including a system selected for the reversal it predicted, indicating that whether compression helps depends on where a system's informative region lies. Compression is therefore better treated as an option than as a default: a linear-only variant is worth reporting alongside the MSS baseline, at the cost of one training run. For a particular system, the choice of compression can be informed by that system's informative region. We examined only the two endpoints of compression, and intermediate degrees are left to future work.

\bibliographystyle{IEEEbib}
\bibliography{refs,mjkim}

\end{document}